# Second generation instruments for the Laser Interferometer Gravitational Wave Observatory (LIGO)

Peter Fritschel

LIGO Project, 175 Albany St., Massachusetts Institute of Technology, Cambridge, MA 02139[1]

## ABSTRACT

The interferometers being planned for second generation LIGO promise an order of magnitude increase in broadband strain sensitivity–with the corresponding cubic increase in detection volume–and an extension of the observation band to lower frequencies. In addition, one of the interferometers may be designed for narrowband performance, giving further improved sensitivity over roughly an octave band above a few hundred Hertz. This article discusses the physics and technology of these new interferometer designs, and presents their projected sensitivity spectra.

## 1. INTRODUCTION

The initial LIGO (Laser Interferometer Gravitational-Wave Observatory) interferometers were designed, as much as possible, using proven concepts and technologies; the idea was to get into the business of collecting and analyzing gravitational wave data quickly, yet to build into the LIGO facilities the capability of supporting much more sensitive instruments in the future. Following the collection of at least one integrated year of science data with the initial detectors, we plan to continue carrying out this strategy by implementing an advanced set of interferometers. The designs being developed promise an improvement over the initial LIGO strain sensitivity by a factor of 15 in the spectral region of maximum sensitivity, 100 Hz < f < 200 Hz, and also reduce the lower end of the sensitive band from 40 Hz to 10 Hz. The impressive effect of these improvements is that the first few hours of LIGO II operation will surpass the space-time volume probed by the entire initial LIGO science run. The strain sensitivity goals for this next generation of LIGO interferometers advance the likelihood of gravitational wave detection from the domain of plausible with initial LIGO, to .

To achieve these sensitivity advances, virtually every aspect of the initial LIGO interferometer design must be upgraded. A new seismic isolation system is needed to push the seismic wall down to 10 Hz; new suspensions and test masses are needed to dramatically reduce thermal noise; much higher laser power and a more efficient interferometer configuration are needed to push down sensing noise; more massive test masses are needed to combat the increased radiation pressure.

Figure 1 shows a schematic diagram of the advanced LIGO interferometer, and Table 1 highlights the performance and design differences between the initial and advanced instruments. Given that we cannot be certain what gravitational wave signals, if any, the initial interferometers will detect, designing the second generation interferometers to have broadband performance is judged to be the most prudent route to 'discovery'. Specific design choices and interferometer parameters are quantitatively evaluated principally against their impact on the sensitivity to neutron star binary inspirals (NBI). In this sense the design is optimized for neutron star binary inspirals, but doing so also tends to give good broadband performance. Design choices are also made with consideration given to technical breakpoints–all feasible attempts are made to improve strain sensitivity, though there may be no significant improvement in the detection sensitivity of known sources. Figure 2 shows the projected sensitivity of the baseline advanced design; the sensitivity is brought very close to the limit imposed by gravity gradients at 10 Hz, but otherwise it is still far from facility limitations, such as that imposed by residual gas in the beam tubes–that challenge is left to yet later generation instruments.

---

1.     email: pf@ligo.mit.edu; http://www.ligo.caltech.edu/

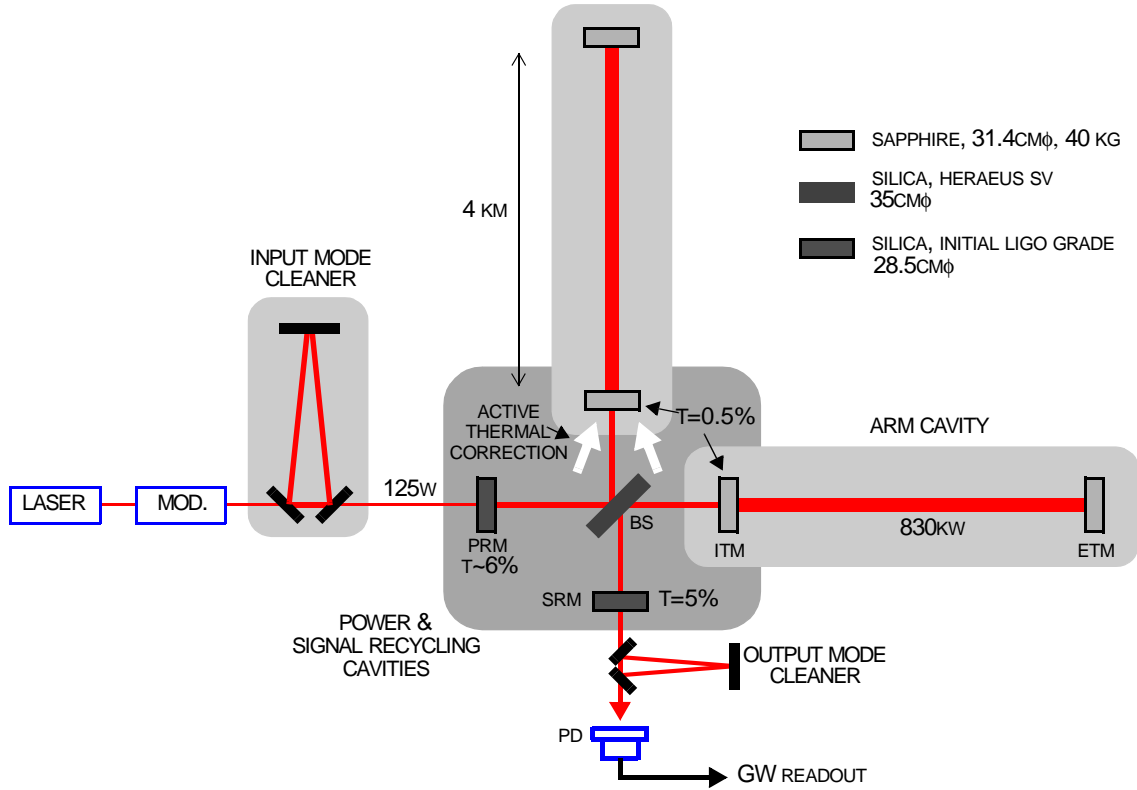

Figure 1: Schematic of an advanced LIGO interferometer, with mirror reflectivities optimized for neutron star binary inspiral detection. Several new features compared to initial LIGO are shown: more massive, sapphire test masses; 20× higher input laser power; signal recycling; active correction of ermal lensing; an output mode cleaner. This is a snapshot of an evolving design. ETM = end test mass; ITM = input test mass; ETM = power recycling mirror; SRM = signal recycling mirror; BS = 50/50 beam splitter; PD = photodetector; MOD = phase modulation.

## 2. INTERFEROMETER CONFIGURATION

The optical configuration of the advanced design (Fig. 1) is a power-recycled and signal-recycled Michelson interferometer, with Fabry-Perot cavities in the arms–thus the basic initial LIGO configuration is kept, and a signal recycling mirror is added at the output[1]. With the initial LIGO configuration, the maximum *response* to a gravitational wave occurs at DC, and we can only control the bandwidth about DC by choice of input test mass transmission (cavity finesse). The additional mirror provides capability to tailor the interferometer response to strain: we can choose the frequency of maximum response and the bandwidth about that point, and use these parameters to optimize the detection sensitivity to a particular source or source class. Figure 3 illustrates the types of possible response curves. The peak response frequency is selected by varying the microscopic position of the signal recycling mirror, although for maximum response at a given frequency the mirror reflectivity must also be optimized. For advanced LIGO, we optimize the response for detection of neutron star binary inspirals; this fixes the signal recycling mirror reflectivity, leaving some limited in-situ freedom to affect the response via the signal mirror's microscopic position. Compared to an optimized interferometer without signal recycling, the additional mirror increases the inspiral detection range by 25-30%, though with a significant technical advantage discussed later.

The diagram in Fig. 3 helps conceptualize the effect of signal recycling. The effect of a passing gravitational wave can be described as producing signal sidebands on the light stored in the arm cavities. Due to the differential character of the gravitational strain, these signal sidebands are anti-phased when they combine at the beamsplitter, and they interfere constructively at the anti-symmetric side of the beamsplitter to travel to the signal recycling mirror. At the signal recycling mirror, the signal sidebands are partially reflected back into the arms; the response of the coupled system is most easily understood by considering that the arm cavity sidebands 'see' a compound output coupler formed by the arm input mirrors and the signal recycling mirror. The equivalent, frequency-dependent reflectivity of this compound mirror may be higher, or lower than the reflectivity of the arm input mirror, depending on the individual mirror reflectivities and the signal frequency. The main laser field is on a Michelson dark fringe and thus does not see the signal recycling mirror; the arm buildup of the laser field is determined only by the optical properties of the arm cavities, the power recycling mirror and the beam splitter. At one end of the design space, the arm cavity bandwidth is made very narrow (high finesse), but is effectively increased for the signal field by the signal recycling mirror, limited by losses in the signal cavity (formed by the signal recycling mirror, the beam splitter, and the arm input mirrors); in this case the laser power in the power cavity (formed by the power recycling mirror, the beam splitter, and the arm input mirrors) is relatively low. At the other end of the design space, the arm cavity bandwidth is made very wide (low finesse), but is effectively narrowed by the signal recycling mirror for the signal field; in this case there is a relatively high buildup in the power cavity, limited again by optical losses. The choice of mirror reflectivities is made considering the impact of the different losses: nonlinear thermal absorption losses in the bulk material argue for lower power recycling gain, and thus higher arm finesse; whereas losses in the signal recycling cavity limit how high the arm finesse can be made and still effectively lowered for the signal field through their coupling into this cavity. Given our estimates of these losses, the optimization between these two effects appears to occur around an arm finesse of 1000. With this, we achieve a single-interferometer NBI range of about 200 Mpc, with 2.1 kW of power on the beamsplitter; without signal recycling, we could attain a range close to this–180 Mpc–but we would need 36 kW of power on the beamsplitter, a level at which thermal absorption would produce devastating distortions.

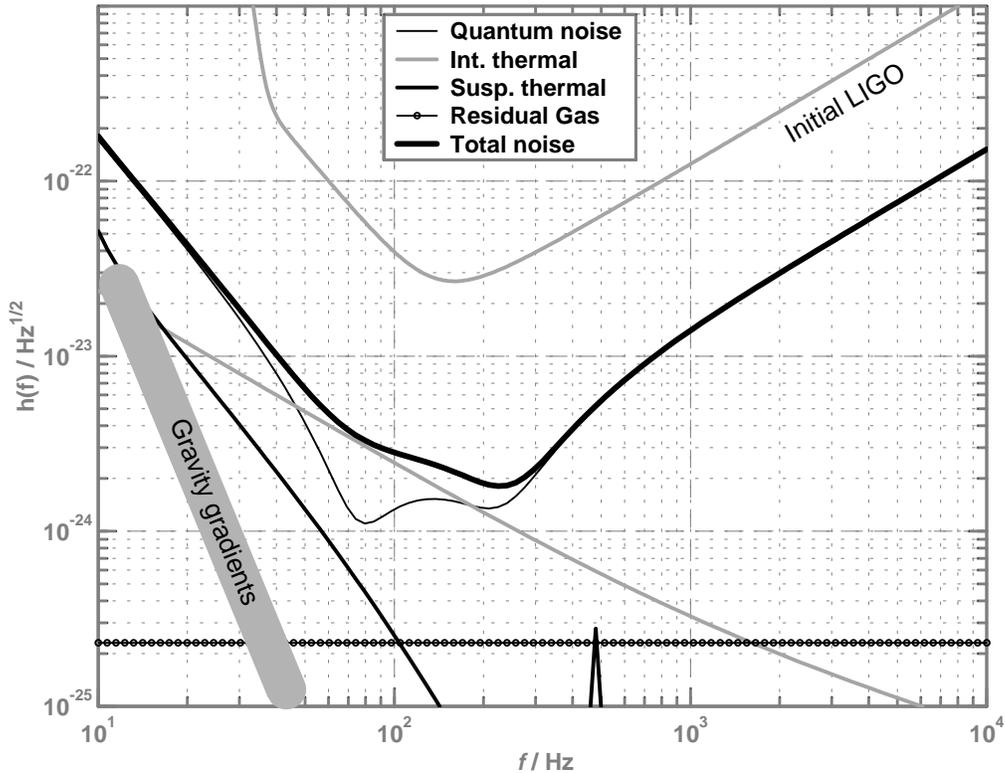

Figure 2: Strain sensitivity projections for advanced LIGO, showing the total noise curve using sapphire test masses, as well as the quantum and thermal noise components for the baseline design. Also shown is the estimate for gravity gradient noise during quiescent times, from reference 2, and the strain noise due to residual gas ($10^{-8}$ torr of $H_2$) in the beam tubes.

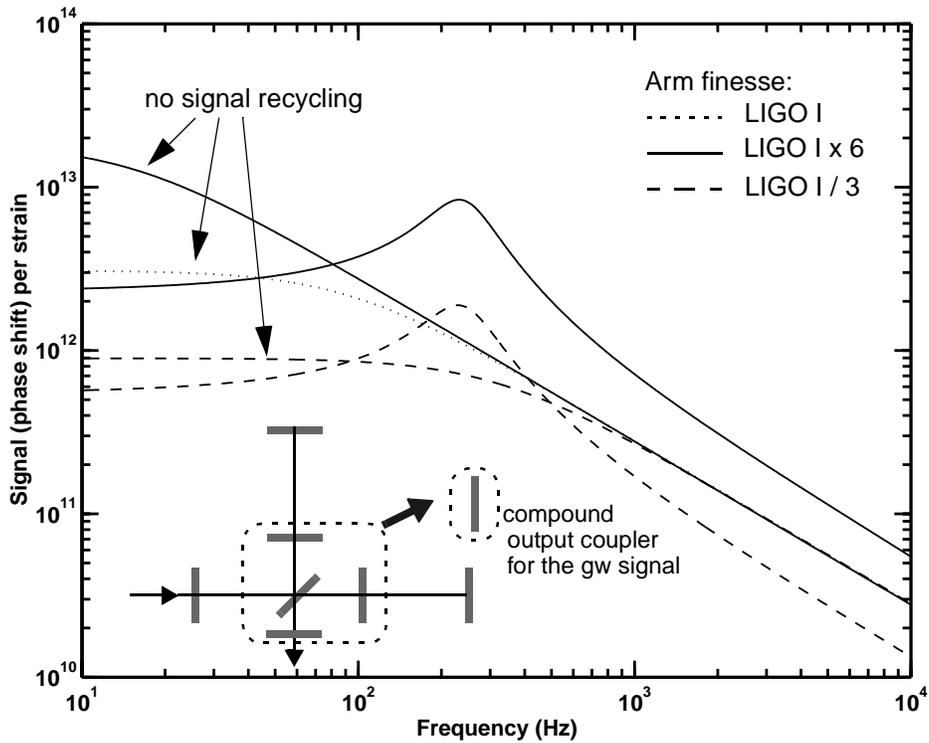

Figure 3: Interferometer gravitational-wave response curves with and without signal recycling. Shown is the optical phase shift per strain for three values of arm finesse: same as LIGO I; 6x the LIGO I finesse; 1/3 of the LIGO I finesse. For the latter two arm finesses, a response curve with signal recycling is shown; the high arm finesse signal recycled curve corresponds to the nominal LIGO II design, and the other curve illustrates that the same response shape can be obtained with a different arm finesse, by adjusting the signal mirror parameters. The low arm finesse signal recycled case has a overall smaller response, but this is compensated by the fact that the beamsplitter power would be higher, in theory yielding the same sensitivity for these two cases. In practice, nonlinear thermal lensing losses favor the design with higher arm finesse and lower beamsplitter power.

Until recently, sensing noise was calculated simply by multiplying the inverse of the response function (given in radians/strain) by the shot noise implied by the power on the beam splitter, given in radians/√Hz. Quantum radiation pressure was then calculated by enforcing the photon number-phase uncertainty relation. Recent work has shown that this oversimplifies the situation. The full quantum mechanical approach shows that the addition of the signal recycling mirror results in a dynamic correlation between quantum shot noise and quantum radiation pressure [3]. At high and low frequencies the noise may still be said to be due to photon counting statistics and radiation pressure, respectively, but at intermediate frequencies–in fact in the most sensitive band–the two effects becomes indistinguishable. Thus in plots of strain sensitivity, where once there were two curves for shot noise and radiation pressure, there is now a single 'quantum noise' curve. The intriguing outcome of the analysis is that over a significant frequency range the free-mass standard quantum limit can be beaten, but at the price of increasing noise at other frequencies.

### 3. THERMAL NOISE

The advanced LIGO design comes close to being a completely quantum noise limited interferometer, with the exception of the 50-250 Hz region, where internal thermal noise of the test masses is dominant. Improved internal thermal noise performance relative to initial LIGO comes from using sapphire for the test mass material. Sapphire has much lower internal frictional losses than fused silica, reflected in a much higher mechanical quality factor–$2 \times 10^8$ versus $2 - 3 \times 10^6$ for a typical initial LIGO test mass. Unfortunately, this advantage is offset somewhat by sapphire's higher thermo-elastic damping[4]. In this noise mechanism, thermodynamical temperature fluctuations act through the material's coefficient of ther-mal expansion to produce fluctuations of the test mass surface. Sapphire suffers relatively strongly from this because of its high thermal expansion and thermal conductivity coefficients. The effect can be countered to some degree by increasing the beam size on the mirrors, thereby reducing the temperature gradients in the material and the corresponding heat loss: a larger beam does a better job of averaging over the local surface fluctuations. As a result we have increased the beam size from 3.7 cm and 4.3 cm radius (on the input and end mirrors, respectively, in the initial interferometers) to 6.0 cm radius

| Parameter | Initial LIGO | Advanced LIGO |
|---|---|---|
| Equivalent strain noise, minimum | $3\times10^{-23}/\sqrt{Hz}$ | $2\times10^{-24}/\sqrt{Hz}$ |
| Neutron star binary inspiral detection range[a] | 19 Mpc | 285 Mpc |
| Stochastic background sensitivity[b] | $3\times10^{-6}$ | $1.5-8\times10^{-9}$ |
| Interferometer configuration | Power-recycled Michelson w/ FP arm cavities | Initial LIGO, plus signal recycling |
| Input laser power | 6 W | 125 W |
| Test masses | fused silica, 11 kg | sapphire, 40 kg |
| Suspension system | single pendulum, steel wires | quad pendulum, silica fibers/ribbons |
| Seismic isolation system, type | passive, 4-stage | active, 2-stage |
| Seismic wall frequency | 40 Hz | 10 Hz |

Table 1: Comparison of interferometer design and performance for Initial and Advanced LIGO.

a. Range for all three LIGO interferometers, assuming for advanced LIGO that two are optimized for binary inspirals and the third is a narrowband instrument.
b. The number is $h_{100}^2 \cdot \Omega_{gw}$, where $h_{100}$ is the Hubble constant in units of 100 km/s/Mpc and $\Omega_{gw}$ is the GW energy density per unit logarithmic frequency interval in units of the closure density.

on all mirrors in advanced LIGO. Such a large beam size requires large mirror dimensions that push the limit of what is obtainable in sapphire, a material not as maturely devoloped as fused silica. Though producing large pieces appears to be possible, good optical uniformity may be difficult to achieve. Sapphire also displays much higher optical absorption than silica–less harmful in theory given sapphire's high thermal conductivity, but potentially more problematic if the absorption is spatially nonuniform.

Research and development of sapphire is underway in industry and within the LIGO Science Collaboration to address these issues, but given the uncertainties with sapphire we are also maintaining a design based on fused silica test masses. Silica offers room for improvement compared to initial LIGO. When lossy materials such as magnets (used in initial LIGO to control the mirror position) are kept off the surface, the intrinsic $Q$ of fused silica is seen to be much higher, with $Q \approx 5\times10^7$ recently measured[5]. Increasing the beam size helps here as well, though not as quickly as with thermoelastic damping (the latter being insignificant in silica, since it is much less thermally conductive and expansive than sapphire).

Both test mass types will also suffer to some degree from another source of thermal noise: mechanical loss in the multi-layer dielectric coatings that are deposited to give the required reflectivities. The coatings, the most common being alternating layers of $SiO_2$ and $TaO_5$, tend to be more lossy than the underlying substrate, possibly leading to significantly increased thermal noise [6]. Research is ongoing to determine the loss mechanism(s) at work in the coatings, and to develop alternative coating materials and/or techniques that will produce low mechanical loss, as well as low optical loss.

In Figure 4 a comparison is shown of the thermal noise for sapphire and fused silica test massees, both for the bare material and for the estimated effects of the dielectric coatings. The slopes of the curves are different because the materials are dominated by different sources of loss: sapphire is dominated by thermo-elastic loss, while fused silica is dominated by internal frictional loss. Clearly, the mechanical loss of the coating will have to be kept below $\sim 10^{-4}$ to preserve the intrinsic loss of the bulk materials.

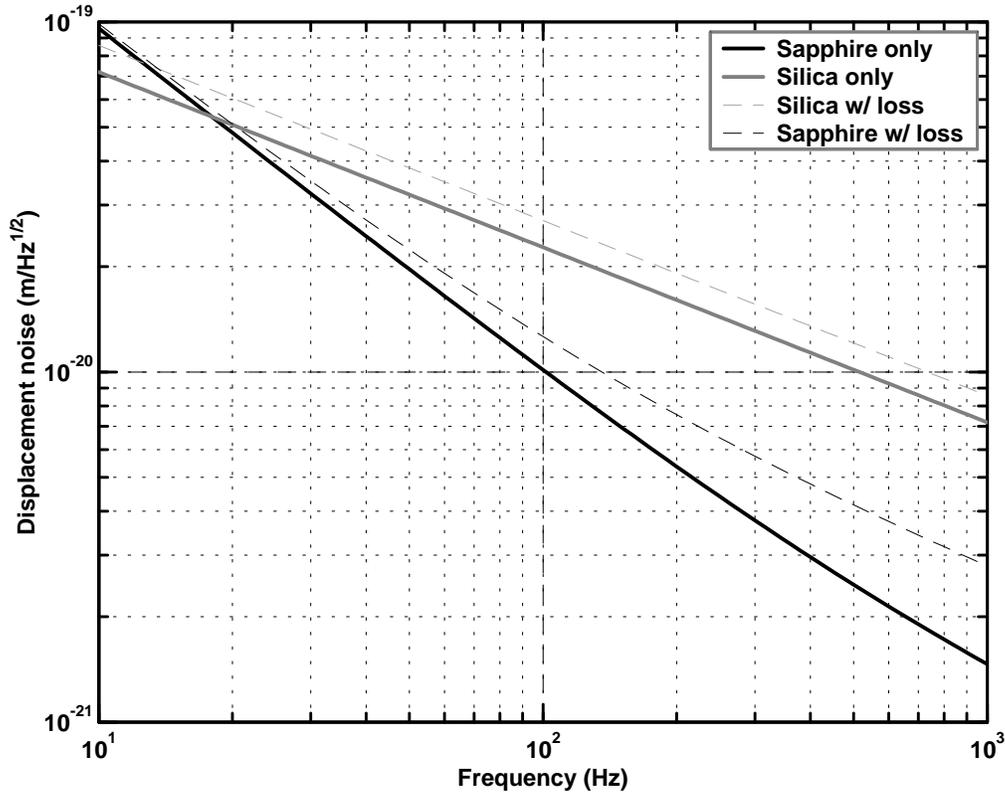

Figure 4: Internal thermal noise for an advanced LIGO test mass, for the nominal sapphire test mass design and for a potential fused silica design. For each material, the solid line shows the thermal noise arising from the bulk material alone, and the dashed line shows the thermal noise prediction when a dielectric coating loss of $10^{-4}$ is added in. In each case the total mass is 40 kg, and the aspect ratio is optimized for lowest noise. The beam size is 6.0 cm radius for the sapphire case, and 5.5 cm radius for the silica case. A coating loss of $10^{-4}$ is at the low end of what has been observed to date[6].

For either material, the test mass size is increased significantly for advanced LIGO. Simply increasing the mass from 10 kg to 40 kg is important to counteract the low-frequency regime of quantum noise, where radiation pressure is dominant. Larger diameter is needed to support the increased beam size in the cavities. We are aiming to use the largest test masses obtainable with the requisite quality, though we are still far from the 1-ton test mass often touted for the ultimate interferometer design.

Thermal noise in the mirrors' suspension systems has also been greatly reduced in the advanced LIGO design. This is primarily the result of using fused silica fibers to suspend the test mass, which exhibit approximately $10^4 \times$ lower intrinsic loss than the steel wires used in initial LIGO[7]. The fiber ends must be attached carefully to the test mass and the stage above–the mechanical joints must not produce additional loss. This is accomplished through the new technique of hydroxy-catalysis bonding[8]; this is used to bond a small interface block to the test mass, and the fiber is then welded to the interface block. The suspension thermal noise can be further reduced by tailoring the cross-sectional geometry of the fiber. For example, for the same level of stress in the fiber, an appropriately oriented ribbon (width much greater than the thickness) will be more compliant along the suspended mass's sensing direction, further diluting the intrinsic loss of the fiber. Another possibility involves choosing the cross sectional area of the fiber such that the linear thermal expansion is cancelled by the Young's modulus temperature dependence, nulling out thermo-elastic damping. These advances have made suspension thermal noise–one of the dominant noise sources for initial LIGO–essentially insignificant for advanced LIGO.

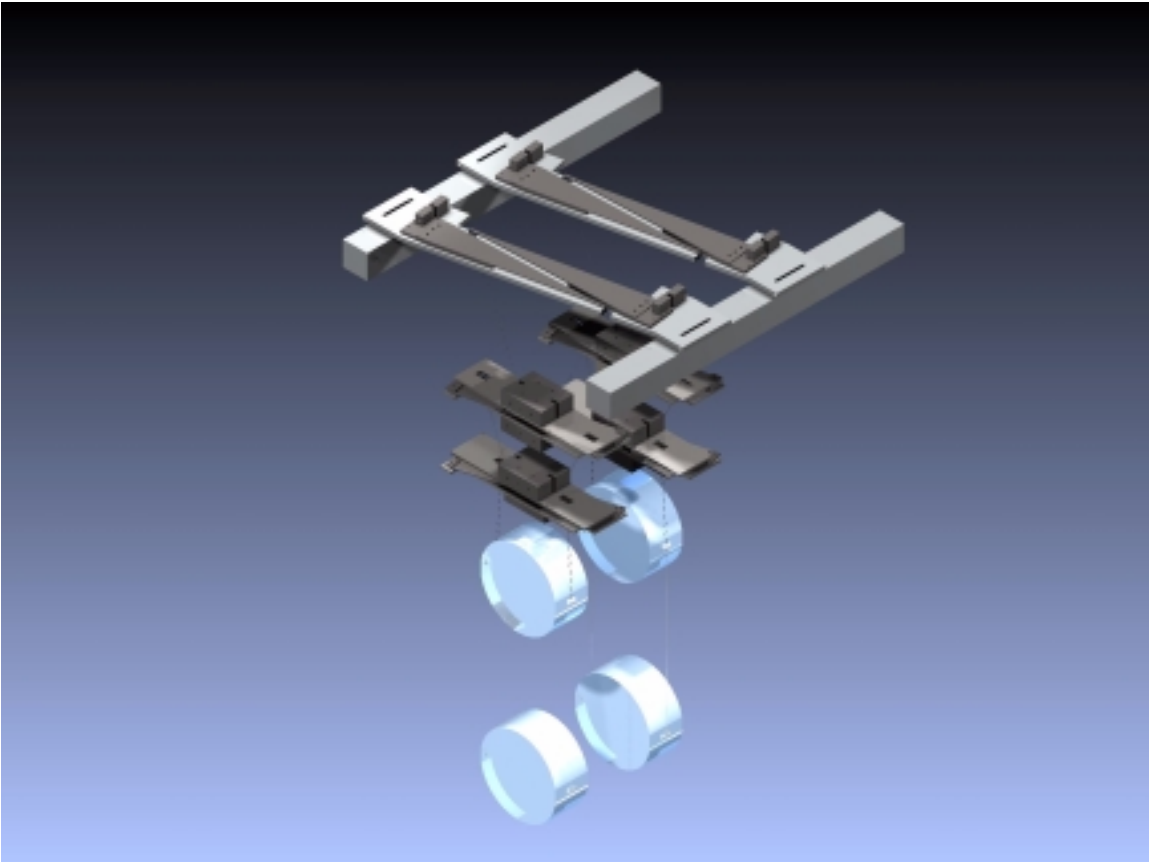

Figure 5: Quadruple pendulum for the advanced LIGO test masses. The bottom mass is the test mass optic, and is suspended from the penultimate mass by fused silica fibers (either circular or ribbon geometry). The upper stages use cantilevered blade springs for high vertical compliance. All local damping is applied at the upper-most suspended stage.

Providing sufficient damping of the rigid body modes of such a low-noise pendulum, without introducing excess noise, is a major design challenge. The solution is to use a multiple-stage pendulum, and damp all modes from the top stage so that the noise in the damping controls can be filtered by the lower stages. A four-stage suspension is needed for sufficient filtering (Fig. 5); it is an extension of the triple pendulums used in the GEO 600 interferometer. In addition to the optic axis isolation provided by the four stages, much greater vertical isolation is achieved through the use of cantilevered blade springs for high vertical compliance.

## 4. SEISMIC NOISE

The four suspension stages naturally provide a great deal of seismic noise attenuation as well–a factor of $10^7$ at 10 Hz. Additional attenuation is given by the seismic isolation system, which is a two-stage active isolation system [9], a large departure from initial LIGO's 4-stage passive system. This system uses a collection of high-sensitivity seismometers and geophones to sense and stabilize via feedback all rigid body degrees-of-freedom of its two stages. The isolation it provides is complementary to the suspension isolation, giving an attenuation of $10^3$–$10^4$ from 1-10 Hz, with significant attenuation down to 0.1 Hz. Isolation at these frequencies below the gravitational wave band is crucial for controlling various technical noise sources, such as laser amplitude noise and noise in the interferometer's global control system.

The heavy filtering of seismic noise provided by the suspensions and seismic isolation create a 'seismic wall' in frequency space–at the test masses, seismic noise falls roughly as $1/f^9$ around 10 Hz. This creates a seismic cutoff frequency, $f_c$, where seismic noise moves the test masses by the same amount as the predominant fundamental noise source, either quantum noise (radiation pressure) or thermal noise. The seismic cutoff frequency goal for advanced LIGO is 10 Hz, pushed down from 40 Hz for initial LIGO. This goal is based partly on consideration of specific astrophysical source detection, and partly on technical feasibility. The estimated neutron star binary inspiral range does not in fact change significantly for $f_c$ below ~20 Hz, though detection of more speculative black hole-black hole mergers could benefit from $f_c < 20$ Hz [10]. Technically, a cutoff at approximately 10 Hz appears feasible without undo risks in the design, so we choose to preserve strain sensitivity down to about 10 Hz. A technical issue at these frequencies is the highest vertical eigen-frequency of the suspension, corresponding to the mode where the fused silica fibers are stretching. It may be difficult to push this eigen-frequency below 10 Hz, due to limitations on fiber stress, fiber length, and other factors. So this vertical mode may be as high as 12 Hz, and its thermal excitation would rise above the noise background (due to quantum radiation pressure) over a 1-2 Hz bandwidth. Even so, this does not necessarily exclude gravitational wave observations in this band. A promising Kalman filtering technique[11], developed to remove higher frequency modes of the suspension fibers, may be applied to the data stream to remove this randomly driven oscillator.

## 5. HIGH LASER POWER

The 20-fold increase in laser power for advanced LIGO is the result of continued incremental progress in laser technology. It will continue to be a diode-pumped Nd:YAG laser as in initial LIGO, and the wavelength will remain at 1064 nm. The design may be a master oscillator-power amplifier type, as in initial LIGO, or a high-power oscillator injection locked to a stable, low-power master. The higher power raises many technical issues regarding power dissipation in the system. This is most troublesome in the beam splitter and arm cavity input mirrors, where bulk or surface absorption leads to thermal lensing and elastic deformation losses. Since these losses scale with the laser power, the effect on interferometer performance is nonlinear, such that for given absorption coefficients there is a relatively hard upper limit to the power sustainable in the system. The LIGO I interferometer is already operating very near this upper limit. High power is of course needed in the arm cavities–where some fraction may be absorbed by the mirror surfaces–to achieve low quantum sensing noise, but the signal recycling configuration allows us to lower the beam splitter power (and thus the power in the input mirror substrates) to mitigate bulk absorption. We also plan on providing active compensation of the laser beam absorption effects. The idea is to reduce the thermal gradients induced via the main beam heating by adding heat preferentially around the outer volume of the optic, either through a circular radiative heating element positioned close to the optic, or through an external laser beam than is scanned across the optic surface[12]. Such a technique should reduce the optical path distortions by an order of magnitude or more.

## 6. INTERFEROMETER OPTIONS

As Fig. 2 shows, the interferometer optimized for neutron star binary inspiral detection displays good broadband performance as well[1], simply because the inspiral signal is relatively broadband. While such an interferometer could be tuned to higher frequency by adjusting the signal mirror position, it would not have very good performance without changing the signal mirror reflectivity. To provide better high frequency sensitivity, an option being considered is to design the third interferometer for narrowband performance. This is achieved with a higher signal mirror reflectivity; as Fig. 5 shows, an appropriate reflectivity value can still yield good narrowband performance over roughly an octave band, e.g. 500-1000 Hz.

---

1. If we define the bandwidth as delimited by the frequencies where the strain noise is 10× the minimum noise level, the bandwidth for the sapphire design is 30 Hz – 1.5 kHz.

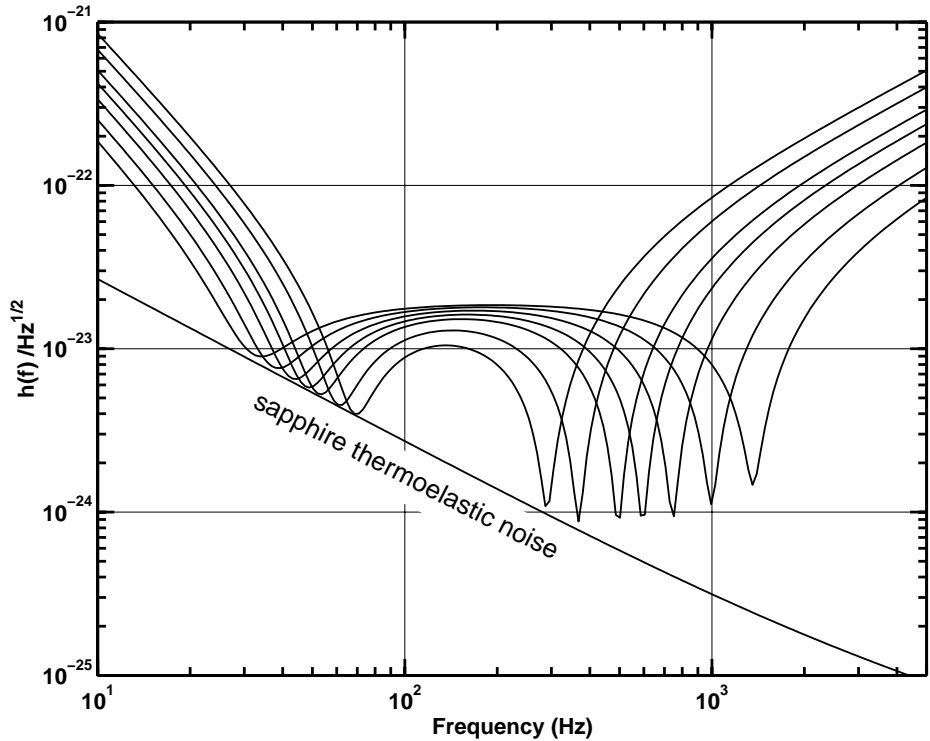

Figure 5: Strain sensitivity curves for a narrowband interferometer. With a single signal recycling mirror chosen to give optimum performance around 700 Hz, good performance between ~500–1000 Hz can be achieved by tuning the signal mirror position microscopically; the set of curves shown span a mirror motion of about $10^{-2}$ wavelength. At the lower end of the octave, sapphire's thermoelastic noise limits the performance; at higher frequencies, above ~500 Hz, sapphire has a clear advantage over fused silica for narrowband performance.

Thus the current strategy is to upgrade all three LIGO interferometers, optimizing two for NS-NS inspiral detection (one at each site), with the third interferometer–increased in length from 2km to 4km–made a tunable, narrowband instrument. An option we will have with the inspiral-optimized detectors is to simply reduce the input laser power. As Fig. 2 shows, the sensitivity will be limited by quantum radiation pressure for frequencies below ~60 Hz. If we reduce the input power to 20-30 W, this low-frequency noise will be reduced as the square root of the power, at the expense of increased noise above ~50 Hz. In this low-power mode of operation, the noise will be limited for $f < 60$ Hz by test mass internal and suspension thermal noise, and gravity gradient noise around 10 Hz. This offers a factor of 2-4 better strain sensitivity at these frequencies, a very attractive improvement for low frequency sources such as a stochastic background, or black hole-black hole coalescences.

The plan is to begin the upgrades at the start of 2006, implementing first one complete interferometer before starting the other two in parallel. I have tried to present here a description (with uncertainties!) of a design which will surely evolve with time, but will continue to be motivated both by what is technically feasible and by astrophysical benchmarks.

## ACKNOWLEDGEMENTS

The second generation LIGO interferometer designs derive from the work of the entire LIGO Laboratory and the LIGO Science Collaboration, all of whom I thank for their contributions. The author is supported by National Science Foundation grant PHY-9210038.

# REFERENCES

1. B.J. Meers, Phys. Rev. D **38**, 2317-2326 (1988).
2. S.A. Hughes and K.S. Thorne, Phys. Rev. D **58** (1998).
3. A. Buonanno and Y. Chen, Phys. Rev. D **64** (2001).
4. V.B. Braginsky, M.L. Gorodetscky, and S.P. Vyatchanin, Phys. Lett. A **264**, 1 (1999); Y.T. Liu, K.S. Thorne, Phys. Rev. D **62** (2000).
5. S.D. Penn, G.M. Harry, A.M. Gretarsson, S.E. Kittelberger, P.R. Saulson, J.J. Schiller, J.R. Smith, S.O. Swords, gr-qc/0009035.
6. G.M. Harry, A.M. Gretarsson, P.R. Saulson, S.D. Penn, W.J. Startin, S. Kittelberger, D. Crooks, J. Hough, G. Cagnoli, N. Nakagawa, S. Rowan, M. Fejer, Class. Quant. Grav. **19,** 897-918 (2002).
7. A.M. Gretarsson, S. Rowan, G. Cagnoli, G.M. Harry, J. Hough, S.D. Penn, P.R. Saulson, W.J. Startin, Phys. Lett. A **A270**, 108-114 (2000).
8. D.H. Gwo, Proc. SPIE-Intl. Soc. Opt. Eng. **3435,** 136 (1998); S. Rowan, S.M. Twyford, J. Hough, D.H. Gwo, R. Route, Phys. Lett. A **246,** 471-478 (1998).
9. S.J. Richman, J.A. Giaime, D.B. Newell, R.T. Stebbins, P.L. Bender, J.E. Faller, Rev. Sci. Inst. **69:6**, 2531 (1998).
10. K. Thorne, private communication.
11. L.S. Finn, S. Mukherjee, Phys. Rev. D **63** (2001).
12. R. Lawrence, M. Zucker, P. Fritschel, P. Marfuta, and D. Shoemaker, Class. Quantum Grav. **19,** 1803-1812 (2002).